\documentclass[12pt]{article}
\usepackage{epsfig}
\usepackage{color}

\newcommand{\be}{\begin{equation}}
\newcommand{\ee}{\end{equation}}

\newcommand{\bea}{\begin{eqnarray}}
\newcommand{\eea}{\end{eqnarray}}
\newcommand{\beq}{\begin{equation}}
\newcommand{\eeq}{\end{equation}}
\newcommand{\nn}{\nonumber}

\def\fun#1#2{\lower3.6pt\vbox{\baselineskip0pt\lineskip.9pt
\ialign{$\mathsurround=0pt#1\hfil##\hfil$\crcr#2\crcr\sim\crcr}}}

\begin{document}

\title{  Hadron diffractive production
 at ultrahigh energies
}
\author{
V.V. Anisovich$^+$, M.A. Matveev$^+$, V.A.  Nikonov$^{+ \diamondsuit}$
}


\maketitle

\begin{center}
{\it
$^+$National Research Centre "Kurchatov Institute",
Petersburg Nuclear Physics Institute, Gatchina, 188300, Russia}

{\it $^\diamondsuit$
Helmholtz-Institut f\"ur Strahlen- und Kernphysik,
Universit\"at Bonn, Germany}

\end{center}

\begin{abstract}
Diffractive production is considered in the ultrahigh energy
region where pomeron exchange amplitudes are transformed into black
disk ones due to rescattering corrections. The corresponding
corrections in hadron reactions $h_1+h_3\to h_1+h_2+h_3$ with small
momenta transferred
($q^2_{1\to 1}\sim m^2/\ln^2s$, $q^2_{3\to 3}\sim m^2/\ln^2s$)
are calculated in terms of
the $K$-matrix technique modified for ultrahigh energies. Small values
of  the momenta transferred are crucial for introducing equations for
amplitudes.
The three-body equation for hadron diffractive production reaction
 $h_1+h_3\to h_1+h_2+h_3$
is written and solved precisely in the eikonal approach.
In the black disk regime final state scattering processes do not
change the shapes of amplitudes principally but dump amplitudes
in a factor $\sim\frac 14$.
\end{abstract}


\section{Introduction}

Recent data for diffractive production of hadrons \cite{totem,auger}
demonstrate a remarkable phenomenon: an appearance of a black spot
in the impact parameter presentation of the $pp$-scattering amplitude.
So, one may suppose that the black disk picture starts at
$\sqrt s\sim 10-100$ TeV.
The key point is the steady growth of total and elastic cross
sections up to the region $\sqrt s\sim 5-50$ TeV, for the preLHC
data see \cite{pre}.
The phenomenon of the increase of high energy cross sections
was discussed for a long time.
First, the power growth, $s^\alpha$ with
$\alpha>1$, was suggested \cite{kaid,DL}
on the basis of the reggeon exchange notion.
Then, it was shown in \cite{Gaisser,Block,Fletcher} that the
power-type growth of scattering amplitudes with energy is dumped
to $\ln^2 s$-type within the $s$-channel unitarization.
The black disk picture at ultrahigh energies is
realized in the Dakhno-Nikonov model \cite{DN} for $\pi p$ and
$p^\pm p$ collisions.
The model, being QCD-motivated, takes
into account the quark structure of colliding
hadrons, the gluon origin of the input pomeron and the colour
screening effects in collisions. The model can be considered as a
realization of the Good-Walker eikonal approach \cite{GW} for a
continuous set of channels.

An appropriate way for the description of the diffractive
scattering data at ultrahigh energies seems to be the use of the profile function in a version of the Good-Walker approach with
the Froissart bound \cite{Froi} (though exceeding the Froissart
bound does not violate necessarily
the general constraints \cite{azimov}).
Examples of such descriptions can be found in
\cite{1110.1479,1111.4984,1201.6298,1202.2016,1208.4086}.

The description of the recent data in terms of the Dakhno-Nikonov model
and the extension of results into the ultrahigh energy region was
performed in \cite{ann1,ann2}, a short summary is given in \cite{annn}.
The fit tells that the 5-50 TeV region turns out to be that
where the asymptotic behaviour starts; the asymptotic regime should
reveal itself definitely at $10^2-10^4$ TeV.

For the ultrahigh energy limit the black disk picture predicts
a $(\ln^2s)$-growth for total and elastic hadron-hadron cross
sections: $\sigma_{tot}\sim \ln^2s$
with
 $[\sigma_{el}/\sigma_{tot}]_{\ln s\to\infty}\to 1/2 $.
Further, the
differential elastic cross sections depend asymptotically on
transverse momenta with a relation for $\tau$-scaling:
 $d\sigma_{el}(\tau)/d\tau= D(\tau )$ with
 $\int_0^\infty d\tau D(\tau)=\sigma_{el}(s)$ and
 $\tau={\bf q}_\perp^2\sigma_{tot}\propto {\bf q}_\perp^2\ln^2s \,$ .
The universal behaviour of all total and elastic
cross sections is the consequence of the universality
of the colliding disk structure, or the structure of parton clouds
at ultrahigh energy. The diffractive dissociation processes are
increasing at asymptotic energies
($\sigma_{D}\propto\ln{s}$, $\sigma_{DD}\propto\ln{s}$)
but their relative contribution tends to zero
($\sigma_{D}/\sigma_{tot}\to 0$, $\sigma_{DD}/\sigma_{tot}\to 0$).

The universal character of the black disks and the $\tau$-scaling
phenomenon open a path for consideration of hadron productions in
diffractive collisions. The simplest process of this type is
the diffractive production of hadron shower, $p+p\to M_X+p$,
the cross section of this process at moderately high energies is
usually modeled by three-pomeron diagrams. More complicated for
consideration is the process of diffractive scattering with the
production of a third particle, $pp\to php$, with large pair
energies ($s_{ph}\sim s_{hp}\sim m\sqrt{s}$) and small momenta transferred to protons (${\bf q}_{p\perp}^2\sim m^2/ \ln^2s$).
This process is the subject of our studies in this paper (see
also Fig. \ref{23f-3}).

The eikonal approach for the black disk picture means a composite
structure of a colliding object: a standard example is the
interaction of a fast particle with a nucleus when multiple elastic scatterings on nucleons of a nucleus give $\eta\to 0$ for the $hA$-amplitude. Time ordering of scatterings is a necessary step in the consideration of such processes, and it results in the eikonal approach. Partons, being hadron components, form the inner
structure of hadrons thus justifying the use of the eikonal
approach for hadron collisions.

Using the $K$-matrix method, or the dispersion relation N/D-approach,
we get an appropriate way for the consideration of three-particle
production processes \cite{book4}. Therefore, the first problem we
face here is the extension of these techniques to ultrahigh energies.
In Chapter 2 we consider examples of diagrams for the
production of three hadrons with very small momenta transferred,
$q^2\sim m^2/\ln^2s $.
The examples demonstrate us specific features in the formulation of
the eikonal approach with the $K$-matrix.
In Chapter 3 in the framework of the developed technique we write in the impact parameter space a system of equations which determines the diffractive production amplitude in hadron-hadron collisions at ultrahigh energies; we solve the equation. The hypothesis about the black disk structure of the two-particle interaction amplitude allows an easy calculation of screening effects inherent to ultrahigh energies, results of such calculations are presented in Chapter 4.

In the Conclusion we summarize the results.

\section{Three particle production amplitude
 and initial/final state interactions}

At ultrahigh energies the initial state and final state rescatterings
are to be taken into account: the growth of total and elastic cross
sections definitely tells us that the effect of rescatterings is not small.
Here we consider examples of such processes using the impact parameter representation. But first we recall the corresponding presentation for
the elastic scattering amplitude.

For the two-particle scattering amplitude the standard determination
of the profile function $T(b,\xi)$ in the impact parameter space,
${\bf b}$, can be written as:
\bea \label{23-13}
&&
4\pi\frac{d\sigma_{el}}{dq^2_\perp}=
A^2(q^2_\perp,\xi),\quad
A(q^2_\perp,\xi)=\int d^2b\; e^{i{\bf b}{\bf q}_\perp} T(b,\xi)\,,
\nn \\
&&
T(b,\xi)=1-e^{-\frac12\chi(b,\xi)}
=1-\eta(b,\xi)\, e^{2i\delta(b,\xi)}=
\frac{-2iK(b,\xi)}{1-iK(b,\xi)}, \nn \\
&& b=|{\bf b}|, \quad \xi=\ln s,
\nn
\eea
where the profile function is presented in terms of the optical density
$\chi(b,\xi)$, the inelasticity parameter and the phase shift, $\eta(b,\xi)$ and $\delta(b,\xi)$,
and using the $K$-matrix approach (the function $K(b,\xi)$ for
the multichannel case is complex valued).

Below we calculate explicitly examples of diagrams for the
production of three particles using the $K$-matrix technique in
the ${\bf b}$-space.

\subsection{
Feynman diagram technique and eikonal approach}

\begin{figure}
\centerline{
\epsfig{file=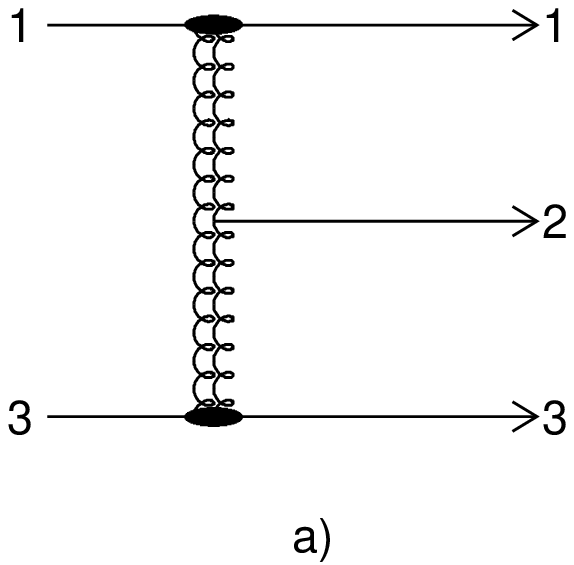,height=4.cm}\hspace{1.5cm}
\epsfig{file=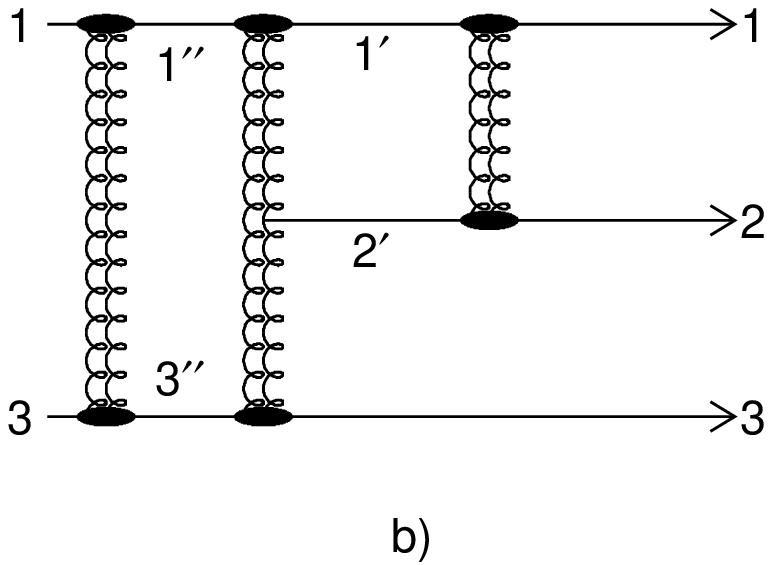,height=4.cm}}
\centerline{
\epsfig{file=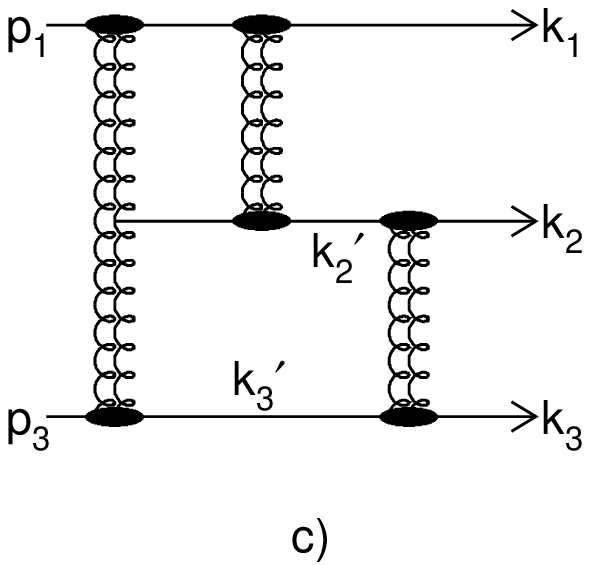,height=4.cm}\hspace{1.5cm}
\epsfig{file=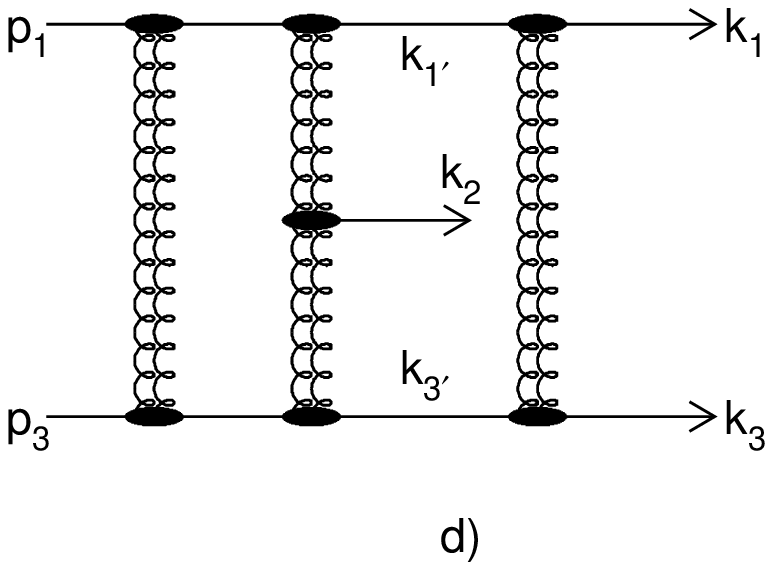,height=4.cm}}
\caption{Input diagram for diffractive production (a) and diagrams
with rescatterings in initial and final states:
 the last interactions in 12-channel (b), in 23-channel (c) and
13-channel (d).
\label{23f-3}}
\end{figure}

Let us consider the amplitude of the Fig. \ref{23f-3}c,d type
(last interaction in 23-channel).
The Feynman integral for the loop related to the
intermediate $2'3'$-state reads:
\be \label{23-5}
\int\frac{d^4k_{2'}}{(2\pi)^4i}\;A_{2\to 3}\left(k_1,k_{2'},k_{3'}\right)
\frac{1}{(m^2-k^2_{2'}-i0)(m^2-k^2_{3'}-i0)}\;
A_{2\to 2}\left(k_{2'},k_{3'}\right)\,.
\ee
The key point is that interactions at ultrahigh energies turn out
to be effectively instantaneous.
This is the result of shrinking of the diffractive cones (the effect
of the $\tau$-scaling): the substantial regions of integration over
momenta transferred are small, of the order of
${\bf q}^2\sim m^2/\xi^2$ with $\xi\equiv \ln s>>1.$

It is convenient to consider three-particle production
processes in the cm-system
where the initial particle momenta are determined as
$p_1=(p_0,{\bf p}_\perp,p_z)\simeq(p+m^2/2p,0,p)$ and
$p_3\simeq(p+m^2/2p,0,-p)$.
Therefore, in this system we have the following relations:
\bea \label{23-2}
&&
{\bf k}_{1\perp}+{\bf k}_{2'\perp}+{\bf k}_{3'\perp}=0,\quad
{\bf k}_{1\perp}+{\bf k}_{2\perp}+{\bf k}_{3\perp}=0,
\\
&&
m^2_{2'\perp}=m^2+{\bf k}^2_{2'\perp},\quad
q^2_1=(p_1-k_1)^2\simeq-{\bf k}^2_{1\perp},
\nn
\\
&&
q^2_{3'}=(p_3-k_{3'})^2\simeq-{\bf k}^2_{3'\perp},\quad
q^2_{3'3}=(k_{3'}-k_{3})^2\simeq
-({\bf k}_{3\perp}-{\bf k}_{3'\perp})^2\,.
\nn
\eea
The integral (\ref{23-5}) within this kinematics is written as:
\be \label{23-7}
\int\frac{dk_{2'}^{(+)}dk_{2'}^{(-)}d^2{k}_{2'\perp}}{2i(2\pi)^4}
\;
\frac{A_{2\to 3}\left(k_1,k_{2'},k_{3'}\right)
A_{2\to 2}\left(k_{2'},k_{3'}\right)}{\left(k_{2'}^{(+)}k_{2'}^{(-)}
-(m^2+{\bf k}^2_{2'\perp})+i0\right)
\left(k_{3'}^{(+)}k_{3'}^{(-)}-(m^2+{\bf k}^2_{3'\perp})+i0\right)}
\,.
\ee
where
$k_{2'}^{(\pm)}=k_{2'0}\pm k_{2'z}$ and
$k_{3'}^{(\pm)}=k_{3'0}\pm k_{3'z}$.
The eikonal approach corresponds to the mass-on-shell
calculation of loop diagrams. It can be seen
when considering the $K$-matrix elements.
The $K$-matrix function $(-i)K(b,\xi)$ of a scattering amplitude
is real for the black disk regime.
This means that the imaginary parts in loop diagrams
are dominant. For the rescattering diagrams of the type given
in Fig. \ref{23f-3},
it is realized by the replacement:
\bea
&&
\left[(m^2-k^2_{2'}-i0)(m^2-k^2_{3'}-i0)\right]^{-1}\to
-2\pi^2\delta(m^2-k^2_{2'})\delta(m^2-k^2_{3'}) \\
&=& -2\pi^2
\delta\left(k_{2'}^{(+)}k_{2'}^{(-)}
-(m^2+{\bf k}^2_{2'\perp})\right)
\delta\left(k_{3'}^{(+)}k_{3'}^{(-)}-
(m^2+{\bf k}^2_{3'\perp})\right)\;.
\nn
\eea
Then the amplitude (\ref{23-5}) reads (below we skip the
index $\perp$):
\be
\int\frac{d^2{k}_{2'}}{(2\pi)^2} \frac{i}{4s_{23}}
\left[A_{2\to 3}\left(k_1,k_{2'},k_{3'}\right)
A_{2\to 2}\left(k_{2'},k_{3'}\right)\right]^{k^2_{2'}=m^2}_{k^2_{3'}=m^2}
\; .\nn
\ee
For mass-on-shell amplitudes we introduce the notation:
\bea
&&
\left[A_{2\to 3}\left(k_1,k_{2'},k_{3'}\right)
\right]^{k^2_{2'}=m^2}_{k^2_{3'}=m^2} \equiv
A^{(23)}\left({\bf k}^2_{1},\xi_{12};\;{\bf k}^2_{3'},\xi_{23}\right)\,,
\\
&&
\frac{1}{4s_{23}}\left[
A_{2\to 2}\left(k_{2'},k_{3'}\right)\right]^{k^2_{2'}=m^2}_{k^2_{3'}=m^2}
\equiv
K_{2\to 2}\left(
({\bf k}_{3'}-{\bf k}_{3})^2, \xi_{23} \right)\,,
\nn
\eea
where $K\left( ({\bf k}_{3'}-{\bf k}_{3})^2, \xi_{23} \right)$
is the $K$-matrix function in momentum representation,
 $\xi_{23}=\ln s_{23}$ and $s_{23}=(k_{2}+k_{3})^2$.
After changing integration
$d^2{\bf k}_{2'}\to d^2{\bf k}_{3'}$, we, finally, write for
Eq. (\ref{23-5}):
\be
\label{23=9}
\int\frac{d^2{k}_{3'}}{(2\pi)^2}
A^{(23)}\left({\bf k}^2_{1},\xi_{12};\;{\bf k}^2_{3'},\xi_{23}\right)\;
iK\left(
({\bf k}_{3'}-{\bf k}_{3})^2, \xi_{23} \right)\,.
\ee
The procedure of calculating mass-on-shell rescatterings given
here corresponds exactly to the eikonal approach used in \cite{DN,ann1,ann2}.
The Fourier transform of Eq. (\ref{23=9})
(the operator reads as
$\int d^2{k}_1/(2\pi)^2   \exp{\left(-i{\bf k}_1{\bf b}_1 \right)}
 \int d^2{k}_3/(2\pi)^2
\exp{\left( -i{\bf k}_3{\bf b}_3 \right)}$)
gives us the amplitude in the impact parameter space.

\subsection{Amplitudes of initial and final state rescatterings in the
 impact parameter space}

We continue to consider examples of diagrams with initial
and final state screenings but in terms of the $K$-matrix. The
best way for that is to use the impact parameter representation.
For the scattering amplitude we write
$T(b,\xi)=\left(-2i K(b,\xi)\right)/\left(1-i K(b,\xi)\right)$,
for production amplitudes we should use the corresponding
Fourier transforms.

\subsubsection{Initial state rescatterings}

The bare amplitude for the production of three particles
and its Fourier transform
(see Fig. \ref{23f-3}a) are written as:
\be
\phi_{0}({\bf k}^2_{1},\xi_{12};\;{\bf k}^2_{3},\xi_{23})
=\int d^2{b}_{1}d^2{b}_{3}f_{0}(b_{1},\xi_{12}\;
;b_{3}\,,\xi_{23})
\exp\left(i{\bf k}_{1}{\bf b}_{1}+i{\bf k}_{3}{\bf b}_{3}
\right).
\label{08-9}
\ee
We use the same notations for the bare amplitude and
its Fourier transform supposing that do not lead to
misunderstanding.

The rescattering in the initial state gives an additional
factor in the impact parameter space
\bea
\phi_{1}({\bf k}^2_{1},\xi_{12};\;{\bf k}^2_{3},\xi_{23})
&=&\int d^2{b}_{1}d^2{b}_{3}
\; iK(b,\xi)f_{0}(b_{1},\xi_{12}\;;b_{3}\,,\xi_{23})
\exp\left(i{\bf k}_{1}{\bf b}_{1}+i{\bf k}_{3}{\bf b}_{3}
\right),
\nn
\\
\xi&=&\xi_{12}+\xi_{23},\quad {\bf b}={\bf b}_{1}+{\bf b}_{3},
\label{08-10}
\eea
two rescatterings result in
 $\left[iK(b,\xi)\right]^2$
and so on. The summation of all terms
$\sum\limits_{n=0,1,2,...}\phi_n$ generates the standard $K$-matrix
factor
$\left[1-iK(b,\xi)\right]^{-1}$,
and we write for the input term corrected by taking into account
the initial state interactions:
\be  \label{08-13}
\phi({\bf k}^2_{1},\xi_{12};\;{\bf k}^2_{3},\xi_{23})
=\int d^2{b}_{1}d^2{b}_{3}
\; \frac{1}{1-iK(b,\xi)}
f_0(b_{1},\xi_{12}\;;b_{3}\,,\xi_{23})
\exp\left(i{\bf k}_{1}{\bf b}_{1}+i{\bf k}_{3}{\bf b}_{3}
\right),
\ee
Below we use the notation
\be  \label{08-14}
f(b_{1},\xi_{12}\;;b_{3}\,,\xi_{23})=
 \frac{1}{1-iK(b,\xi)}
f_0(b_{1},\xi_{12}\;;b_{3}\,,\xi_{23})
\ee
The factor $\left[1-iK(b,\xi)\right]^{-1}$ is universal for all
terms of the amplitude.
Hence, it should be introduced into all components of the
amplitude. Below, without special emphasizes, we presume that
it is done.

\subsubsection{Input term $
\phi({\bf k}^2_{1},\xi_{12};\;{\bf k}^2_{3},\xi_{23})
$
and final state rescatterings }

First, we consider  rescatterings in 23- and 12-channels. The input
term with the corresponding rescatterings
can be written as:
\bea
&&
\phi^{(23)}({\bf k}^2_{1},\xi_{12};\;{\bf k}^2_{3},\xi_{23})
=\int d^2{b}_{1}d^2{b}_{3}f^{(23)}(b_{1},\xi_{12}\;;b_{3}\,,\xi_{23})
\exp\left(i{\bf k}_{1}{\bf b}_{1}+i{\bf k}_{3}{\bf b}_{3}
\right),\nn
\\
&&
\phi^{(12)}({\bf k}^2_{3},\xi_{23};\;{\bf k}^2_{1},\xi_{12})
=\int d^2{b}_{1}d^2{b}_{3}f^{(12)}(b_{3}\,,\xi_{23}\;;b_{1},\xi_{12})
\exp\left(i{\bf k}_{1}{\bf b}_{1}+i{\bf k}_{3}{\bf b}_{3}
\right),
\label{09-15}
\eea
where
\bea
f^{(23)}(b_{1},\xi_{12}\;;b_{3}\,,\xi_{23})&=&
f(b_{1},\xi_{12}\;;b_{3}\,,\xi_{23})
\frac{iK(b_{3},\xi_{23})}{1-iK(b_{3},\xi_{23})}
\nn
\\
&\equiv&f(b_{1},\xi_{12}\;;b_{3}\,,\xi_{23})\;a(b_{3},\xi_{23}),
\nn
\\
f^{(12)}(b_{3}\,,\xi_{23}\;;b_{1},\xi_{12})
&=&
f(b_{1},\xi_{12}\;;b_{3}\,,\xi_{23})
\frac{iK(b_{1},\xi_{12})}{1-iK(b_{1},\xi_{12})}
\nn
\\
&\equiv&
f(b_{1},\xi_{12}\;;b_{3}\,,\xi_{23})\;
a(b_{1},\xi_{12})\,.
\label{09-16}
\eea
Here we introduce a short notation for the two-particle scattering
amplitude.

Rescatterings in the 13-channel give us the input
term $f^{(13)}$. Since here
${\bf k}_{1}+{\bf k}_{3}={\bf k}_{1'}+{\bf k}_{3'}$, it is convenient
to use the equation for rescatterings in a somewhat modified form:
\bea \label{09-17}
\phi^{(13)}({\bf k}^2_{1},\xi_{12};\;{\bf k}^2_{3},\xi_{23})&=&
\int\frac{d^2{k}_{3'}d^2{k}_{1'}}{(2\pi)^2}\;
\delta\left({\bf k}_{1}+{\bf k}_{3}-{\bf k}_{1'}-{\bf k}_{3'}\right)\\
&\times&
\phi ({\bf k}^2_{1'},\xi_{12};\;{\bf k}^2_{3'},\xi_{23})\;
iK\left(({\bf k}_{3'_\perp}-{\bf k}_{3_\perp})^2,\xi_{13} \right)\,.
\nn
\eea
The transformation of this equality into the impact parameter space
leads to
\bea \label{09-18}
f^{(13)}_1(b_{1}\,,\xi_{12}\;;b_{3},\xi_{23}\; ;|{\bf b}_{1}+{\bf b}_{3}|\,,\xi_{13})
&=&
f(b_{1},\xi_{12}\;;b_{3}\,,\xi_{23})\;
iK(|{\bf b}_{1}+{\bf b}_{3}|\,,\xi_{13}),
\nn\\
 \xi_{13}&\simeq&\xi=\xi_{12}+\xi_{23}\,.
\eea
Performing a summation over the complete set of rescattering
diagrams in the 13-channel,
$\sum\limits_{n=1,2,3...}f^{(13)}_n$, we write:
\bea \label{09-19}
f^{(13)}(b_{1}\,,\xi_{12}\;;b_{3},\xi_{23}\; ;|{\bf b}_{1}+{\bf b}_{3}|\,,\xi)
&=&
f(b_{1},\xi_{12}\;;b_{3}\,,\xi_{23})
\frac{iK(|{\bf b}_{1}+{\bf b}_{3}|\,,\xi)}{1-iK(|{\bf b}_{1}+{\bf b}_{3}|\,,\xi)}\\
&\equiv&
f(b_{1},\xi_{12}\;;b_{3}\,,\xi_{23})\;
a\left(|{\bf b}_{1}+{\bf b}_{3}|\,,\xi\right).
\nn
\eea
Formulae (\ref{09-16}), (\ref{09-19}) give us the pattern
for writing other diagrams with final state rescatterings.

\begin{figure}
\centerline{\epsfig{file=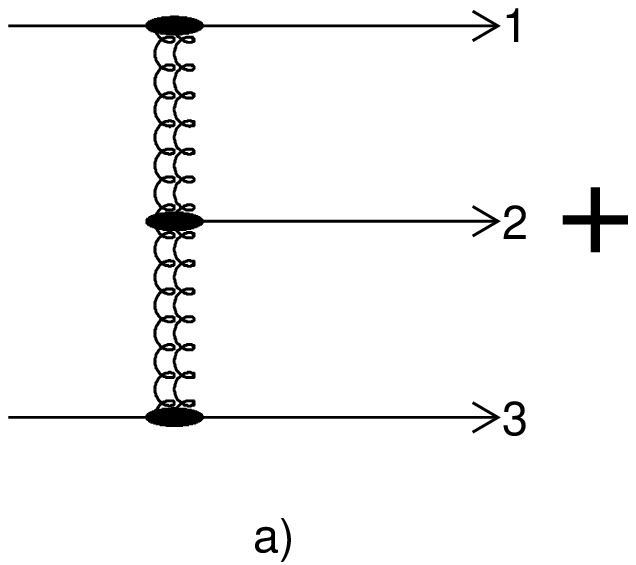,height=30mm}\hspace{-2mm}
            \epsfig{file=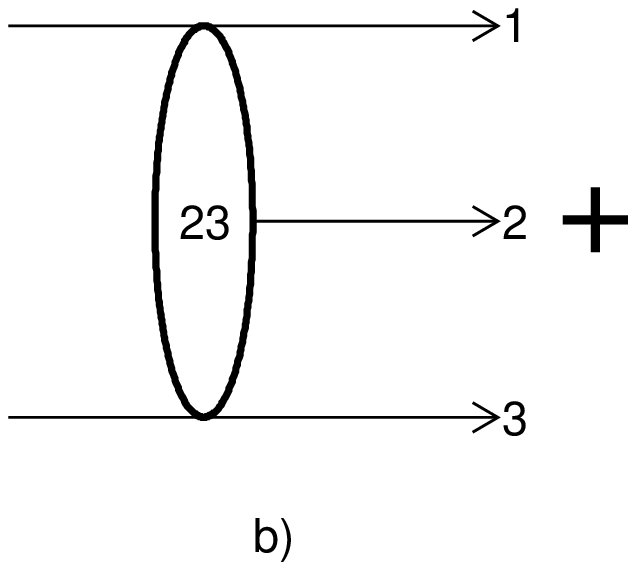,height=30mm}\hspace{-2mm}
            \epsfig{file=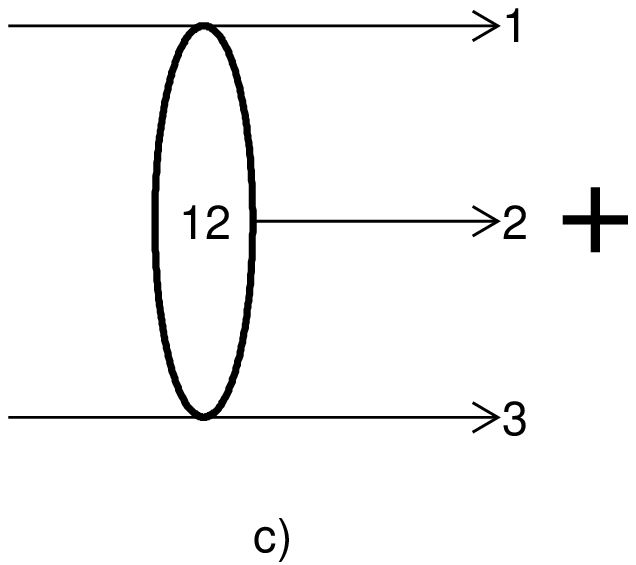,height=30mm}\hspace{-2mm}
            \epsfig{file=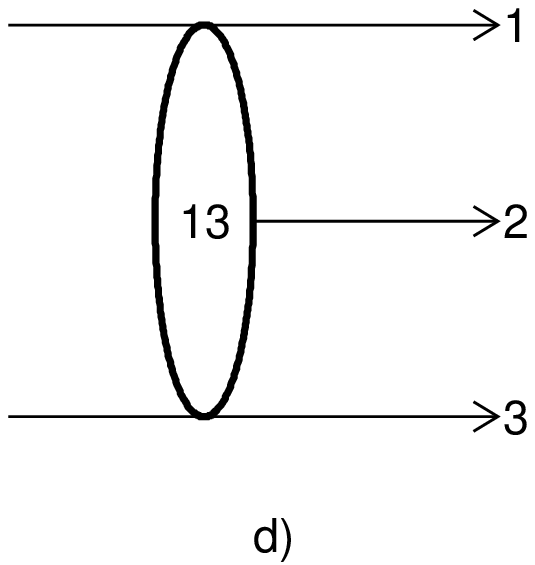,height=30mm}}
\caption{Four terms for a three particle production amplitude: the
input term without final state rescatterings (a), and the terms with
rescatterings in the $23$-channel (b), in the $12$-channel (c),
and the $13$-channel (d).
\label{f-apr2} }
\end{figure}

\section{System of equations for the production amplitude }

\begin{figure}
\centerline{\epsfig{file=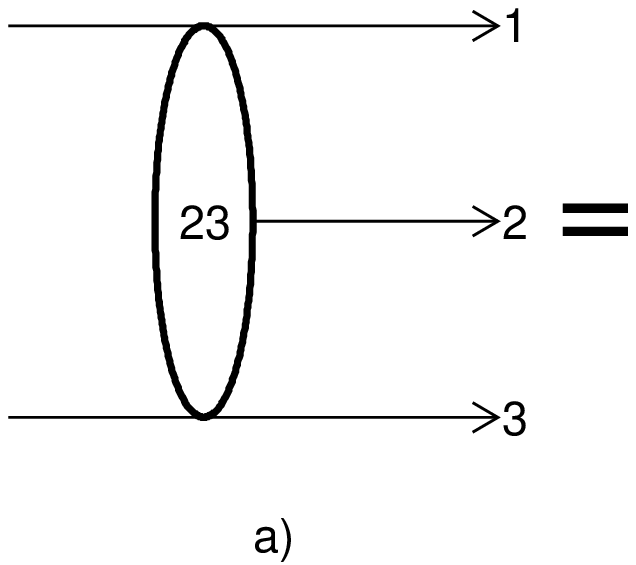,height=30mm}\hspace{-2mm}
            \epsfig{file=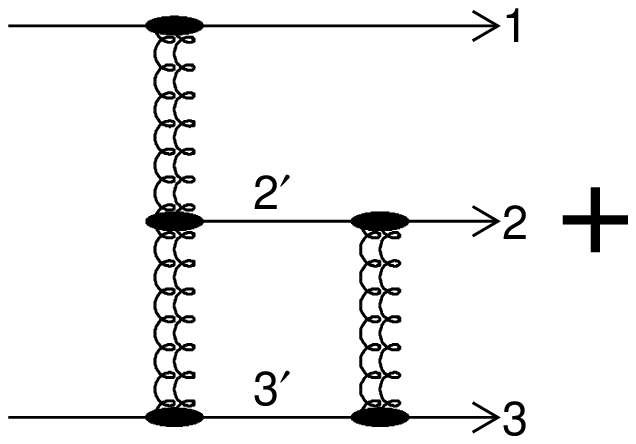,height=30mm}\hspace{-2mm}
            \epsfig{file=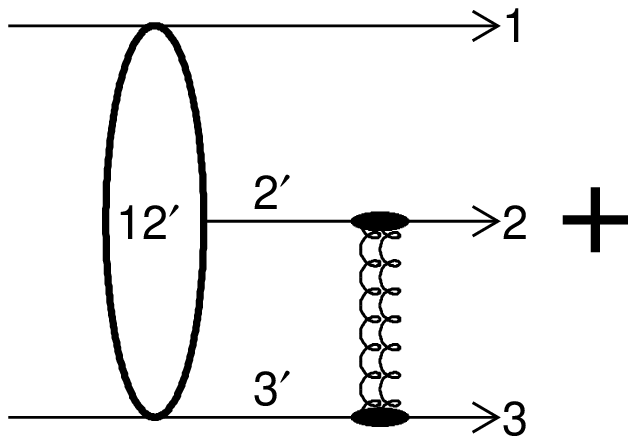,height=30mm}\hspace{-2mm}
            \epsfig{file=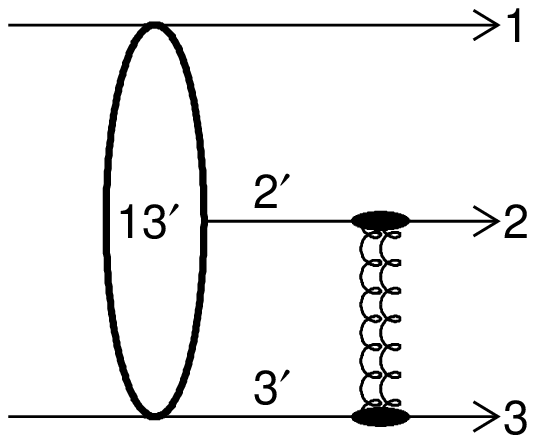,height=30mm}}
\vspace{3mm}
\centerline{\epsfig{file=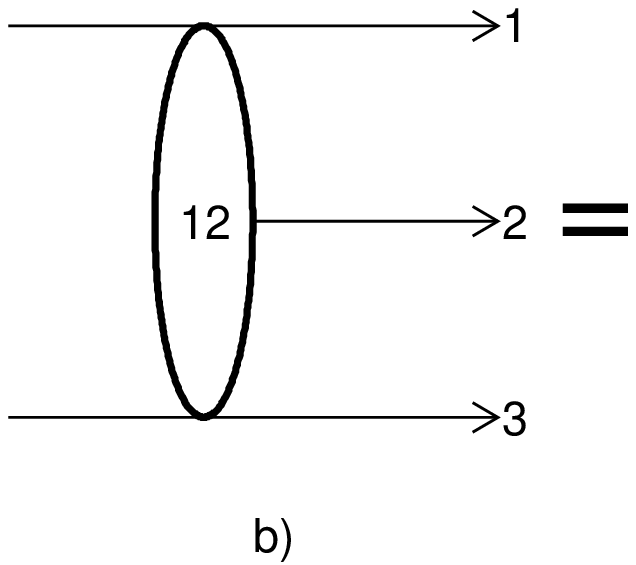,height=30mm}\hspace{-2mm}
            \epsfig{file=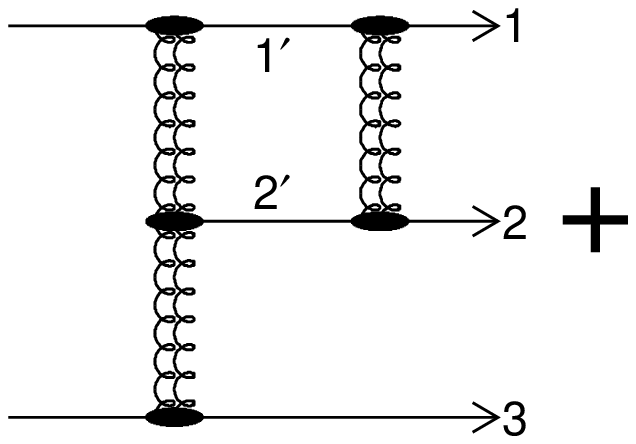,height=30mm}\hspace{-2mm}
            \epsfig{file=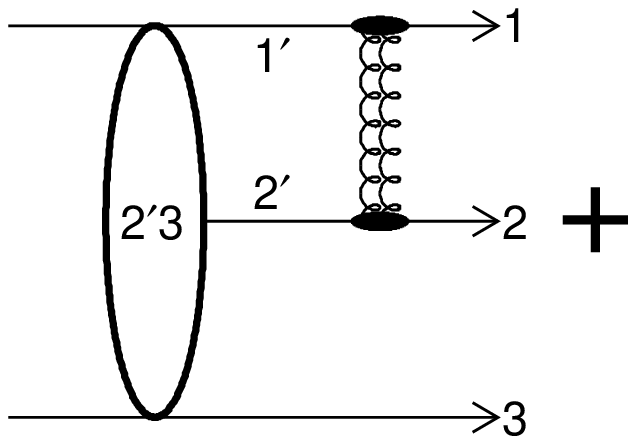,height=30mm}\hspace{-2mm}
            \epsfig{file=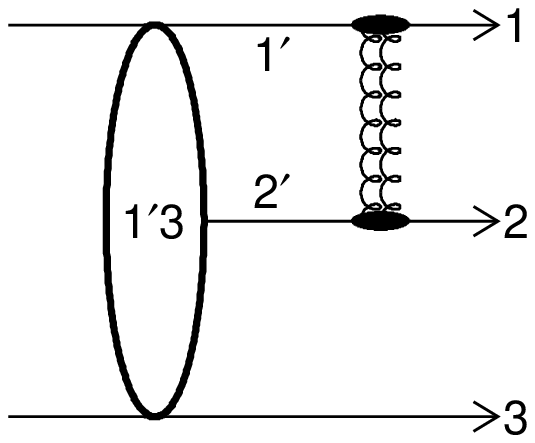,height=30mm}}
\vspace{3mm}
\centerline{\epsfig{file=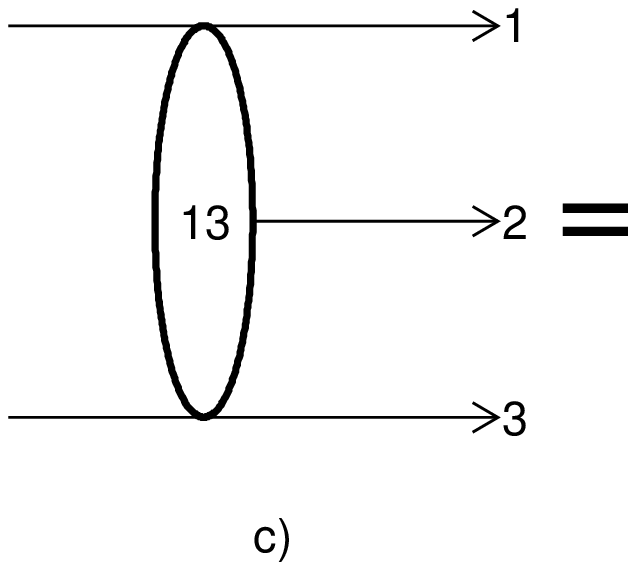,height=30mm}\hspace{-2mm}
            \epsfig{file=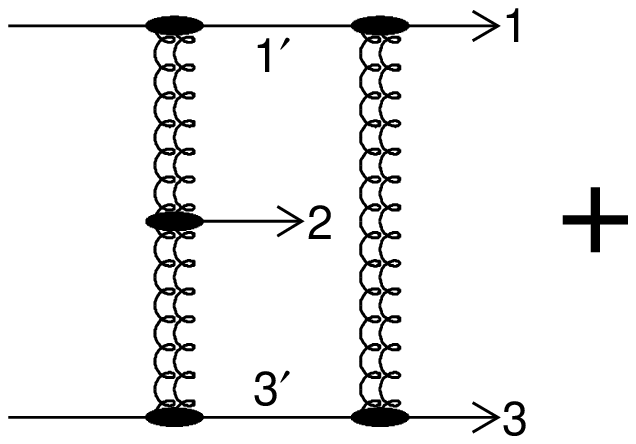,height=30mm}\hspace{-2mm}
            \epsfig{file=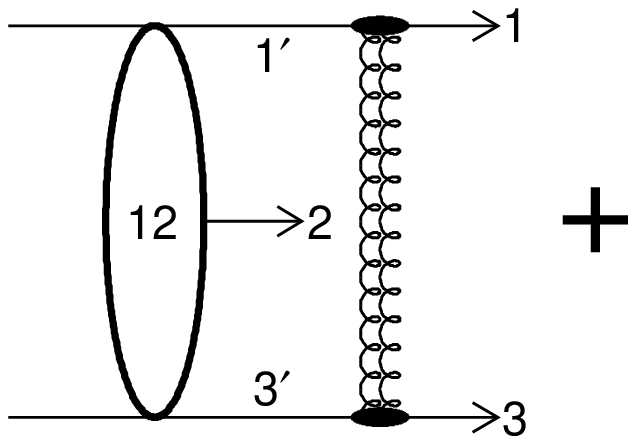,height=30mm}\hspace{-2mm}
            \epsfig{file=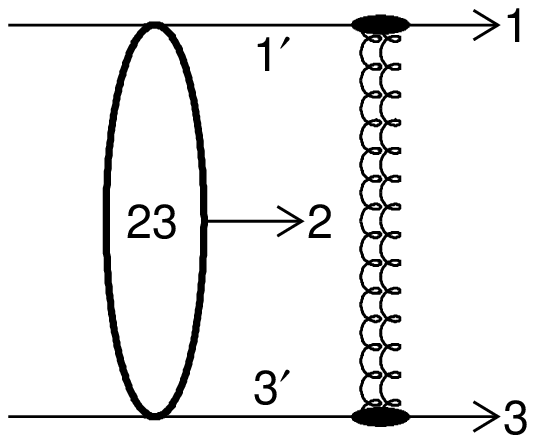,height=30mm}}
\caption{Diagrams with final state rescatterings in the channels
23,12 and 13 - figures $a$, $b$ and $c$ correspondingly. Equalities
are graphical realizations of equations for the amplitudes $A^{(ij)}$.
\label{f-apr3}}
\end{figure}

We present the amplitude as a sum of four terms:\\
(i) input term $f(b_{1},\xi_{12}\;;b_{3}\,,\xi_{23})$ without final
state interactions,  \\
(ii) terms with interactions in final states, that are $12$-,
$23$-states and $13$-state with corresponding amplitudes
$A^{(12)}(b_{1},\xi_{12}\;;b_{3}\,,\xi_{23})$,
$A^{(23)}(b_{1},\xi_{12}\;;b_{3}\,,\xi_{23})$, and
$A^{(13)}(b_{1},\xi_{12}\;;b_{3}\,,\xi_{23};|{\bf b}_{1}+{\bf b}_{3}|\,,\xi)$. \\
The index $(ij)$ shows the channel in which the last interaction
in the final state takes place.
We suppose that initial state interactions in the amplitudes $f$
and $A^{(ij)}$ are taken into account.
Then, the total amplitude, see Fig. \ref{f-apr2}, is written
as follows:
\bea
\label{10-20}
A^{(tot)}_{2\to 3} &=&
f(b_{1},\xi_{12}\;;b_{3}\,,\xi_{23})+
A^{(12)}(b_{1},\xi_{12}\;;b_{3}\,,\xi_{23})
\nn\\
&+& A^{(23)}(b_{1},\xi_{12}\;;b_{3}\,,\xi_{23})+
A^{(13)}(b_{1},\xi_{12}\;;b_{3}\,,\xi_{23};
|{\bf b}_{1}+{\bf b}_{3}|,\,\xi).
\eea
The equations for $A^{(ij)}$ are shown in Fig. \ref{f-apr3},
they are written as:
\bea
&&
A^{(23)}(b_{1},\xi_{12}\;;b_{3}\,,\xi_{23})=
f(b_{1},\xi_{12}\;;b_{3}\,,\xi_{23})\; a_{23}(b_{3},\xi_{23})
\nn
\\
&+&
 A^{(12)}(b_{1},\xi_{12}\;;b_{3}\,,\xi_{23})\; a_{23}(b_{23},\xi_{23})
+A^{(13)}(b_{1},\xi_{12}\;;b_{3}\,,\xi_{23};|{\bf b}_{1}+{\bf b}_{3}|,\,\xi)\; a_{23}(b_{23},\xi_{23}),
\nn
\\
&&
A^{(12)}(b_{1},\xi_{12}\;;b_{3}\,,\xi_{23})=
f(b_{1},\xi_{12}\;;b_{3}\,,\xi_{23})\; a_{12}(b_{1},\xi_{12})
\nn
\\
&+&
 A^{(23)}(b_{1},\xi_{12}\;;b_{3}\,,\xi_{23}) a_{12}(b_{1},\xi_{12})
+A^{(13)}(b_{1},\xi_{12}\;;b_{3}\,,\xi_{23};|{\bf b}_{1}+{\bf b}_{3}|,\,\xi)\;a_{12}(b_{1},\xi_{12})
,
\nn\\
&&
A^{(13)}(b_{1},\xi_{12}\;;b_{3}\,,\xi_{23};|{\bf b}_{1}+{\bf b}_{3}|,\,\xi)
= f(b_{1},\xi_{12}\;;b_{3}\,,\xi_{23})\; a_{13}(|{\bf b}_{1}+{\bf b}_{3}|,\,\xi)
\nn\\
&+&
 A^{(12)}(b_{1},\xi_{12}\;;b_{3}\,,\xi_{23})\;a_{13}(|{\bf b}_{1}+{\bf b}_{3}|,\,\xi)
+A^{(23)}(b_{1},\xi_{12}\;;b_{3}\,,\xi_{23})\;a_{13}(|{\bf b}_{1}+{\bf b}_{3}|,\,\xi)
.
\label{10-21}
\eea
Let us recall that the two-particle amplitudes, $a_{ij}$, are
introduced in Eq. (\ref{09-16}), namely:
\bea
\label{22}
&&
 a_{23}\equiv a_{23}(\xi_{23},b_{23})=
\frac{iK(b_{3},\xi_{23})}{1-iK(b_{3},\xi_{23})}\;,
 \\
&&
a_{12}\equiv a_{12}(b_{1},\xi_{12})=
\frac{iK(b_{1},\xi_{12})}{1-iK(b_{1},\xi_{12})}\;,
\nn\\
&&
a_{13}\equiv a_{13}(|{\bf b}_{1}+{\bf b}_{3}|,\,\xi)=
\frac{iK(|{\bf b}_{1}+{\bf b}_{3}|,\,\xi)}{1-iK(|{\bf b}_{1}+{\bf b}_{3}|,\,\xi)}\;.
\nn
\eea
Amplitudes $A^{(23)}$ and $A^{(12)}$ differ only by the
permutation of indices $1\rightleftharpoons 3$.
In a short form Eq. (\ref{10-21}) reads:
\bea
&&
A^{(23)}=
f a_{23}+A^{(12)}a_{23} +A^{(13)} a_{23},
\nn \\
&&
A^{(12)}=fa_{12}+A^{(23)} a_{12}+ A^{(13)} a_{12},
\nn \\
&&
A^{(13)}= f a_{13}+A^{(12)}a_{13}+A^{(23)}a_{13}.
\label{10-22}
\eea
The equations give us
\bea \label{11-23}
&&
A^{(12)}=
f\frac
{1+a_{13}+a_{23}+a_{23}a_{13} }
{1-a_{12}a_{23}-a_{12}a_{13}-a_{23}a_{13}-2a_{12}a_{23}a_{13}}\,a_{12}
,\nn\\
&&
A^{(23)}=
f\frac
{1+a_{13}+a_{12}+a_{12}a_{13} }
{1-a_{23}a_{12}-a_{23}a_{13}-a_{12}a_{13}-2a_{12}a_{23}a_{13}}\,a_{23}
,\\
&&
A^{(13)}=
f\frac {1+a_{12}+a_{23}+a_{23}a_{12} }
{1-a_{23}a_{12}-a_{23}a_{13}-a_{12}a_{13}-2a_{12}a_{23}a_{13}}
\,a_{13},
\nn
\eea
with
\be
f=\frac{1}{1-iK(b,\xi)}f_0,\qquad
\xi=\xi_{12}+\xi_{23},\quad
{\bf b}={\bf b}_{1}+{\bf b}_{3}\,.
\ee
The input term $f_0$ depends also on ${\bf b}_{j}$ and $\xi_{ij}$.
At moderately high energies it is a two-pomeron term,
at ultrahigh energies it can be a two-disk term. So, for
the two-pomeron term we write
correspondingly in the momentum and impact parameter spaces:
\bea
\label{23_26}
{\rm  {\bf k}-space:}&&\quad f_0=g_{2\to 3}\cdot
ie^{-i\frac{\pi}{2}\Delta}
\exp{\left[\Delta\xi_{12}-\alpha'\xi_{12}{\bf k}^2_1\right]}\cdot
ie^{-i\Delta\frac{\pi}{2}}
exp{\left[\Delta\xi_{23}-\alpha'\xi_{23}{\bf k}^2_3\right]}\,,
\nn\\
{\rm {\bf b}-space:}&&\quad f_0=g_{2\to 3}\cdot
\frac{ie^{-i\frac{\pi}{2}\Delta+\Delta\xi_{12}}}{4\pi\alpha'\xi_{12}}
\exp\left[-\frac{{\bf b}^2_1}{4\alpha'\xi_{12}}\right]
\frac{ie^{-i\frac{\pi}{2}\Delta+\Delta\xi_{23}}}{4\pi\alpha'\xi_{23}}
\exp\left[-\frac{{\bf b}^2_3}{4\alpha'\xi_{23}}\right]\,.
\eea
In the black disk mode the input term reads:
\bea
\label{23_28}
{\rm {\bf k}-space:}&&f_0=g_{2\to 3}\cdot
i A({\bf k}^2_1,\xi_{12})\;i A({\bf k}^2_3,\xi_{23})\,,
\nn\\
{\rm {\bf b}-space:}&&f_0=g_{2\to 3}\cdot
i T(b_1,\xi_{12})\;i T(b_3,\xi_{23})\,
\eea
with $A({\bf k}^2,\xi)$, $T(b,\xi)$ determined in Eq. (\ref{23-13}).

In the black disk mode
the amplitudes $a_{ij}$ behave as $ a_{ij}\to -\frac12$ in the
region $b<R_{black\;disk}\simeq R_0\ln s$
and $a_{ij}\to\; 0$ at large distances (beyond
the black disk area). Therefore the denominator of Eq. (\ref{11-23})
is non-zero, and that results in a unique solution.

\begin{figure}
\centerline{\epsfig{file=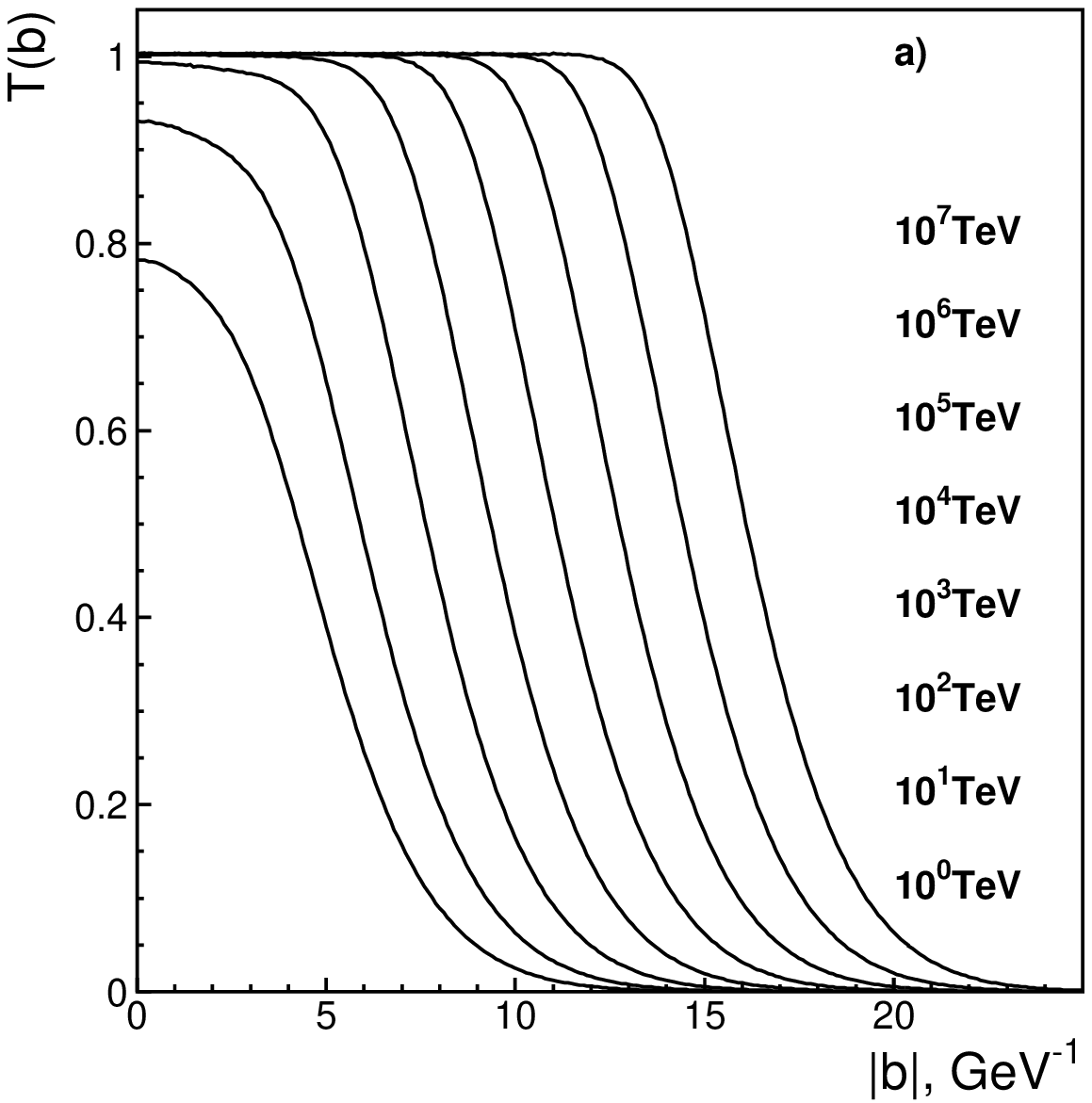,width=8cm}
            \epsfig{file=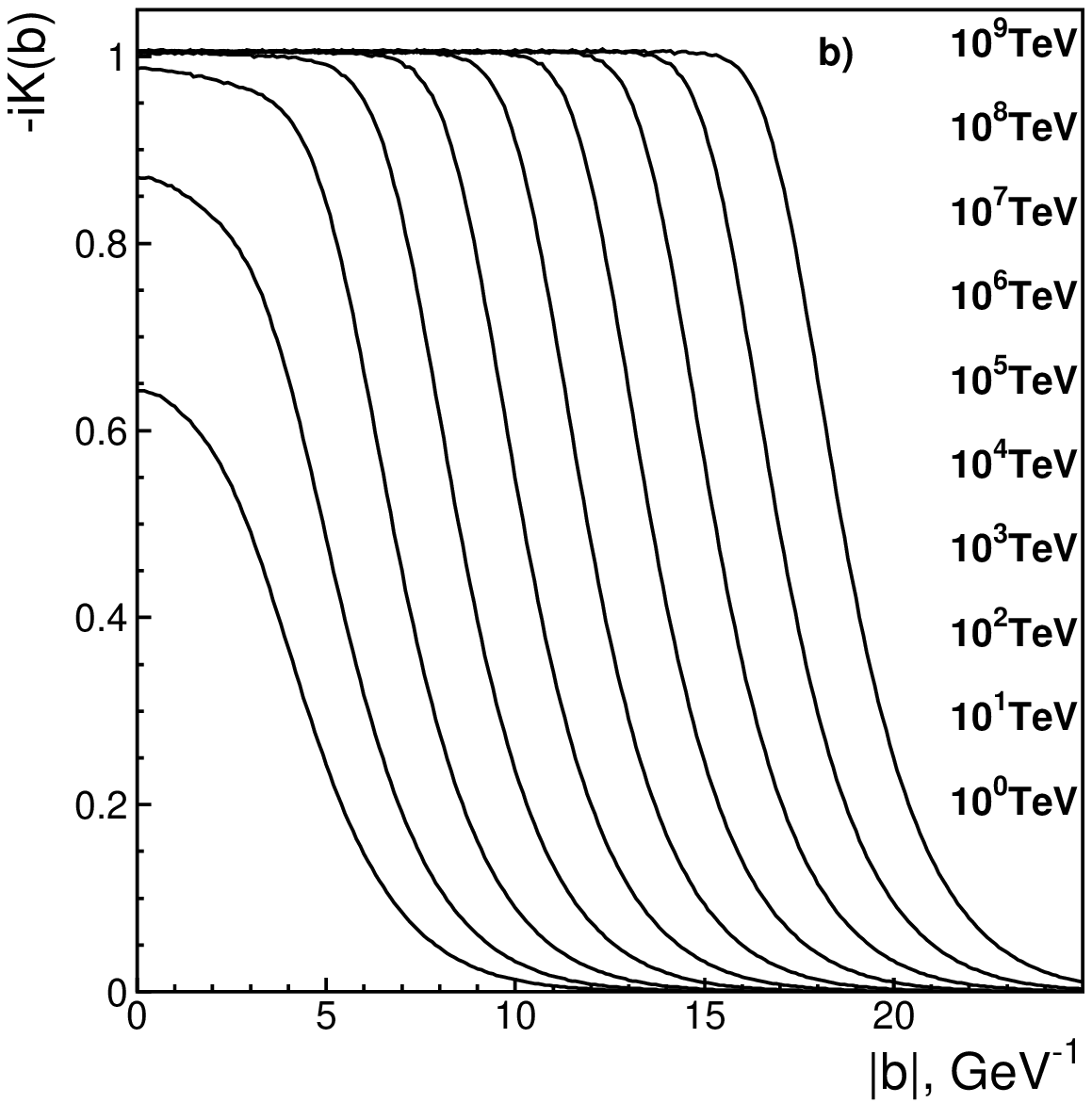,width=8cm}}
\caption{\label{23f-4}
The $K$-matrix function $(-i)K(b,\xi)$ for the $pp$ scattering amplitude at
$\sqrt{s}=1$, $10$ TeV is obtained by the fit of
existing data \cite{totem,auger,pre}. The curves at
$\sqrt{s}=10^n$ TeV at $n\geq 2$ show a continuation
of the fit results \cite{ann1,ann2} to ultrahigh energies in
terms of the black disk picture, here at $b<R_{black\;disk}\simeq R_0\ln s$ one has $(-i)K(b,\xi)\to 1$.}
\end{figure}

\subsection{Black disk mode and numerical solution
of the three-body equation
}

For numerical calculations of $f$, $A^{(ij)}$
and $A^{(tot)}=f+\sum A^{(ij)}$ we use
the Dakhno-Nikonov model \cite{DN} with fit results
for $\left(-iK(b,\xi)\right)$ obtained
in \cite{ann1,ann2}.

\begin{figure}
\centerline{\epsfig{file=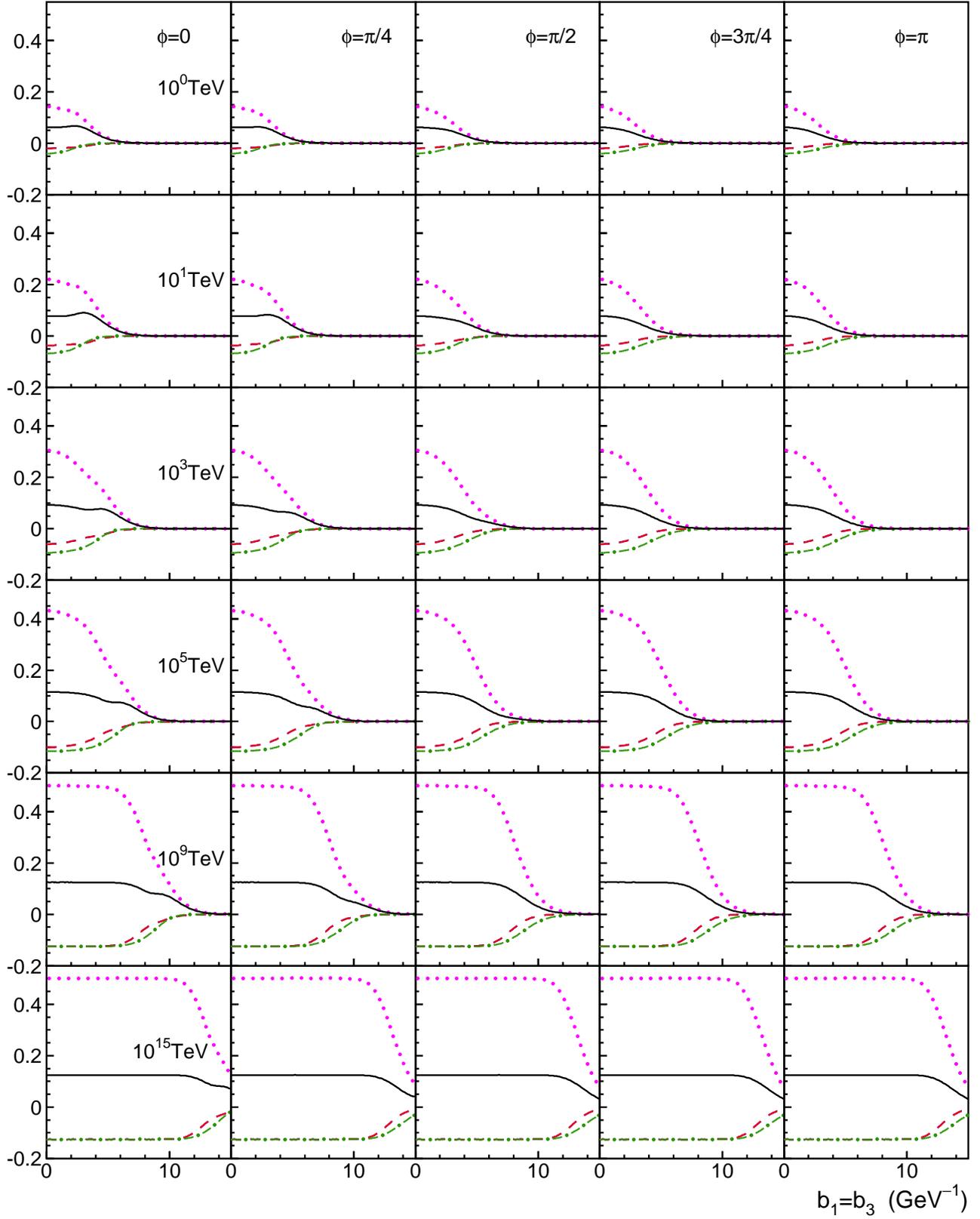,height=0.9\textheight}}
\caption{
Black disk mode: Production amplitudes in impact parameter space for
colliding energies $\sqrt{s}=1,10,10^3,10^5,10^9,10^{15}$ TeV at
$b_1=b_3$ and $\xi_{12}=\xi_{23}=\xi/2$, $\phi$ is the angle between
${\bf b}_{1}$ and ${\bf b}_{3}$: $A^{(12)}=A^{(23)}$ are red dashed
curves, $A^{(13)}$ are green dot-dashed curves, $f$ are pink dotted
curves, $A^{(tot)}$ are black solid curves, see (\ref{10-20}),
(\ref{11-23}), (\ref{23_28}).
 \label{ampl-12}}
\end{figure}

\begin{figure}
\centerline{\epsfig{file=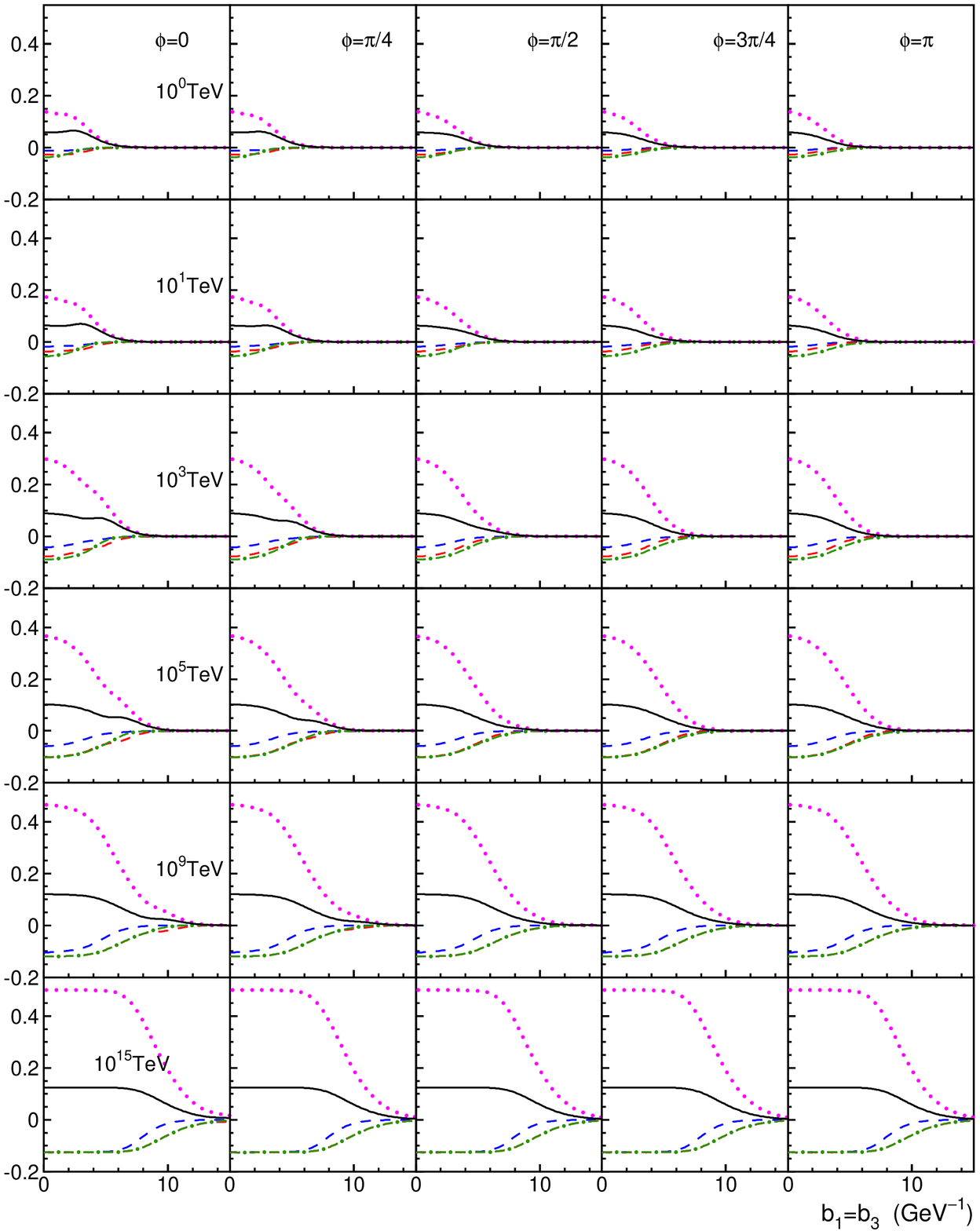,height=0.9\textheight}}
\caption{
Black disk mode: Production amplitudes in impact parameter space for colliding energies
$\sqrt{s}=1,10,10^3,10^5,10^9,10^{15}$ TeV at
$b_1=b_3$ and $\xi_{12}=\xi/3$, $\xi_{23}=2/3\;\xi$,
$\phi$ is the angle
between ${\bf b}_{1}$ and ${\bf b}_{3}$:
$A^{(12)}=A^{(23)}$ are red dashed curves,
$A^{(13)}$ are green dot-dashed curves, $f$ are pink dotted curves,
$A^{(tot)}$ are black solid curves,
see (\ref{10-20}), (\ref{11-23}),
(\ref{23_28}).
\label{ampl-1323}}
\end{figure}

In Fig. \ref{23f-4}a we show the profile functions $T(b,\xi)$ found
in the fit of data \cite{totem,auger,pre}. So, we can regard
the profile functions
at $\sqrt s=1,10,100$ TeV as those restored by the data,
while $T(b,\xi)$ at $\sqrt s=10^{n},\;n>2$ are asymptotic values
for the black disk regime. The corresponding $-iK(b,\xi)$ are shown
in Fig. \ref{23f-4}b. The numerical calculation of $f$,
$A^{ij}$, $A^{tot}$ are performed
using these $\left(-iK(b,\xi)\right)$, the results are demonstrated
in Figs. \ref{ampl-12} and \ref{ampl-1323}.

The comparison of the input amplitude $f$ with $A^{(tot)}$ demonstrates
that final state scattering corrections do not change principally the
shape of $A^{(tot)}$  but dump the amplitude in a factor $\sim\frac 14$.

\section{Conclusion}

The diffractive scattering amplitudes are growing with energy
thus causing the necessity to take into account
rescatterings that result in screening effects.
In this paper we calculate these effects
in the framework of modified $K$-matrix
technique for eikonal amplitudes.
Corresponding calculations of
screening effects in diffractive production processes are
performed in the impact parameter space.

The key point of the approach is a shrinkage of diffractive cones
in hadron reactions at ultrahigh energies. The shrinkage of cones
with energy growth demonstrates us the effective suppression
of the $t$-channel singularities that allow us to use
quasi-instantaneous interactions. Being more detailed, one can suppose
that the long-ranged component of interaction is determined by a
cloud of partons which have universal characteristics
and properties. These long-ranged
interactions are quasi-instantaneous,
thus opening possibilities to a standard
treatment of the multiparticle production processes; examples of
such considerations can be found in \cite{book4}.
The generalization to other ultrahigh energy diffractive
production processes is possible within the developed technique.

The authors thank Y.I. Azimov, J. Nyiri and M.G. Ryskin
for useful discussions and comments.
 The work was supported by grants RFBR-13-02-00425 and
 RSGSS-4801.2012.2.

   \end{document}